\documentclass[preprint,authoryear, 12pt]{elsarticle}
\usepackage{setspace}
\usepackage{color}
\usepackage{lipsum}
\usepackage{amsfonts,amsmath,amssymb,mathrsfs,amsthm}
\usepackage{amsfonts,amsmath,amssymb,mathrsfs}
\usepackage{epstopdf}
\usepackage{natbib}
\usepackage{algorithmic}
\usepackage{amsopn}
\usepackage{mathtools}
\usepackage{cases}
\usepackage{color}
\usepackage{enumerate}
\usepackage{graphicx,subfigure}
\usepackage{hyperref}
\hypersetup{colorlinks=true[linkcolor=false, citecolor=black, urlcolor=black]{hyperref}}
\usepackage{geometry}
\newtheorem{theorem}{Theorem}[section]
\newtheorem{lemma}{Lemma}[section]
\newtheorem{definition}{Definition}[section]
\newtheorem{remark}{Remark}[section]

\newtheorem{proposition}{Proposition}[section]
\newtheorem{corollary}{Corollary}[section]
\newtheorem{example}{Example}[section]

\newcommand{\esssup}{\mathop{\operatorname{ess\,sup}}}

\numberwithin {equation} {section}

\begin{document}
	\journal{arXiv.org}
	\begin{frontmatter}
		\title{Set-valued Star-Shaped Risk Measures}
		
		\address[math]{School of Mathematics, China University of Mining and Technology, Xuzhou, P.R. China}
		
		\author[math]{Bingchu Nie}
		\ead{niebingchu@163.com}

		\author[math]{Dejian Tian\corref{correspondingauthor}}
		\ead{djtian@cumt.edu.cn}
		\cortext[correspondingauthor]{Corresponding author}
		
		\author[math]{Long Jiang}
		\ead{jianglong365@cumt.edu.cn}
		
		\begin{abstract}
			In this paper, we introduce a new class of set-valued risk measures, named set-valued star-shaped risk measures. Motivated by the results of scalar monetary and star-shaped risk measures, this paper investigates the representation theorems in the set-valued framework. It is demonstrated that set-valued risk measures can be represented as the union of a family of set-valued convex risk measures, and set-valued normalized star-shaped risk measures can be represented as the union of a family of set-valued normalized convex risk measures. The link between set-valued risk measures and set-valued star-shaped risk measures is also established.
		\end{abstract}
	\end{frontmatter}
	
	\section{Introduction}\label{sec1}
	
	The use of risk measures to quantify the risk of financial portfolio has been studied extensively in the literature. Initially, an axiomatic description of coherent risk measures satisfying monotonicity, translation invariance, subadditivity, and positive homogeneity was introduced by \cite{1999coherent}. In \cite{2002convex}, and \cite{2002order}, the axioms of coherency have been relaxed to define convex risk measures. The definition of risk measures is generalized to quasi-convex by \cite{2011quasiconvex} and to star-shaped risk measures by \cite{2021starshaped}. In the past two decades, there have been many other developments in various directions; see \cite{2016Stochastic} and the references therein.
	
	In all of the above methods, risk measures are applied to univariate positions with the basic assumption that each asset is evaluated in terms of a num\'{e}raire and the obtained numbers are summed in order to obtain the value of a portfolio. However, these scalar-valued risk measures map a real-valued random variable into a real number, disregarding portfolio aggregation. In markets with frictions, it is more appropriate to evaluate the risk of multivariate random payoffs due to factors such as transaction costs, liquidity bounds, etc. Set-valued risk measures are functions mapping a multivariate random variable into a subset of some finite dimensional space, known as the space of eligible portfolios. The idea is to collect all deterministic initial portfolios which compensate for the risk of a multivariate position into one set and call this set the value of the risk measure at this position. \cite{2004vectorvalued} first introduced the concept of set-valued coherent risk measures. Set-valued convex risk measures and their dual representations have been studied in \cite{2010dualityfor}. Up to this point, research on set-valued risk measures mainly focuses on those satisfying convexity due to the excellent tools of convex analysis. 
 
	On the other hand, in the scalar case, it is widely recognized that convexity is a useful property of acceptance sets and risk measures and many results were obtained in the existing literature. By representing the general risk measures with convex risk measures, the study of general risk measures can be transformed into the study of convex risk measures. \cite{1999coherent} showed that Value at Risk (VaR) can be represented as a lower envelope of a family of coherent risk measures. Recently, \cite{2020riakaversion} provided properties and representation theorems for risk measures satisfying only translation invariance and second-order stochastic dominance using the Expected Shortfall model. On this basis, \cite{2020monetary} proved that a monetary risk measure without convexity can be represented as a lower envelope of a family of convex risk measures. \cite{2021starshaped} investigated a representation of risk measures with star-shapedness as minima of convex risk measures. \cite{2022montaryandstarshaped} established the relationship between the above monetary risk measure and star-shaped risk measure by clarifying the importance of the acceptability of zero. Representation results of cash-subadditive risk measures were studied in \cite{han2021cashsubadditive}.
	
	This paper aims to study star-shapedness of set-valued risk measures. We characterize the class of set-valued star-shaped risk measures by providing properties and representation of its members.
	
	In this paper, first, we will introduce a new class of set-valued risk measures named set-valued star-shaped risk measures. Non-convex risk measures exhibit broader applicability as they impose less stringent requirements, albeit at the cost of losing access to several powerful tools. As a class of non-convex risk measures, the investigation of set-valued star-shaped risk measures aims to relax the convexity conditions of set-valued convex risk measures and extend star-shapedness from scalar-valued to set-valued area. This expansion in scope enables the application of these measures to a wide range of scenarios from two different perspectives, thereby enhancing their versatility across various risk measures. We will  investigate some fundamental properties of star-shapedness, including the relationship between positive homogeneity, convexity and star-shapedness, the equivalent form of star-shapedness, and one-to-one correspondences between star-shaped risk measures and star-shaped acceptance sets. 
	
	Second, we will consider the set-valued extension of representation theorem for monetary risk measures and star-shaped normalized risk measures. By leveraging the one-to-one correspondence between risk measures and acceptance sets, our research on risk measures can be transformed into an examination of acceptable sets. We will demonstrate that set-valued risk measures can be represented by a family of set-valued convex risk measures, while set-valued normalized star-shaped risk measures can be represented by a family of set-valued normalized convex risk measures. The representation form and the construction of acceptance sets are different from the scalar-valued case due to the peculiarities of set-valued risk measures. These theorems serve as a bridge between more general set-valued risk measures and set-valued convex risk measures, allowing research on general set-valued risk measures to be translated into research on convex risk measures and enabling the use of tools from existing convex risk measures.
	
	Third, we establish the relationship between set-valued monetary risk measures and set-valued star-shaped risk measures. Under appropriate conditions, we will demonstrate that set-valued risk measures can be regarded as a translation applied to a specific position within set-valued star-shaped risk measures. The scalar-valued results on risk measures in \cite{2022montaryandstarshaped} will not always hold (see Remark \ref{remark5.2}) in set-valued situation.
	
	We end the introduction with a review of relevant literatures. Set-valued generalizations of some well-known scalar coherent risk measures have already been studied, such as set-valued average value at risk in \cite{2013averagevar}, set-valued shortfall and divergence risk measure in \cite{2017shortfall}, set-valued loss-based risk measures in \cite{2018lossbased} and set-valued cash sub-additive risk measures in \cite{2019cashsub-additive}. \cite{2011setvaluedconical} extended set-valued risk measures to the case of random exchange rates at terminal time. Set-valued risk measures were extended to a dynamic framework in \cite{2013timeconsistency,2015multiportfolio} and \cite{2018discreteprocess,2020dynamicprocess} introduced set-valued risk measures for processes. The relation between set-valued risk measures for processes and that for vectors was proved in \cite{2022process+vectors}. \cite{2020capital} introduced the capital allocation problem in the set-valued context. In addition, \cite{2023dynamicstar}, and \cite{2023dynamicstarwai} have extended the results of scalar-valued star-shaped risk measures to the dynamic domain. \cite{2023law-star} presents characterizations of law-invariant star-shaped functionals. 
	
	The paper is structured as follows. Section \ref{sec2} contains a mathematical model of the situation, including definitions and relationships of set-valued risk measures and acceptance sets. In Section \ref{sec3}, we will state the definition of star-shapedness and provide some propositions of set-valued star-shaped risk measures and star-shaped acceptance sets. In Section \ref{sec4}, we will provide the representation theorems for the set-valued risk measures and set-valued normalized star-shaped risk measures. In Section \ref{sec5}, we establish the transformation relationship between set-valued risk measures and set-valued star-shaped risk measures.  Section \ref{sec6} concludes the paper.

	\section{Preliminaries}\label{sec2}
	
	In this section, we introduce some notations and definitions about set-valued risk measures and its acceptance sets.  The readers can refer to \cite{2010dualityfor} for more related details. 
	
	\subsection{Notations}
	Let $(\Omega, \mathcal{F}, P)$ be a probability space. By $L_d^p=L_d^p(\Omega, \mathcal{F}, P), 1\leq p\leq \infty$, we denote the linear space of all $P$-measurable functions $X: \Omega\to\mathbb{R}^d$ such that $\int_{\Omega}|X|^p dP<\infty$ for $1\leq p<\infty$ and ess.sup$_{\omega\in \Omega}|X(\omega)|<\infty$ for $p=\infty$.  In all cases, $|\cdot|$ denotes Euclidean norm on $\mathbb{R}^d$, and the usual identification of functions differing only on sets of $P$-measure zero is assumed; hence $L_d^p$ is a Banach space for $1<p<\infty$.  
	
An element $X\in L_d^p$ has components $X_1,\cdots, X_d$ in $L^p:=L_1^p$. For any $X=(X^1,\cdots,X^d)$, $Y =(Y^1,\cdots,Y^d)\in L_d^p$, $X+Y$ stands for $(X^1+Y^1,\cdots, X^d+Y^d)$ and $aX$ stands for $(aX^1,\cdots,aX^d)$ for $a\in \mathbb{R}$. For sets $A$ and $B$ in $L_d^p$, $A+B=\{X+Y:X\in A, Y\in B\}$, and $\emptyset+B=B+\emptyset=\emptyset$.
	
	Let $K\subseteq \mathbb{R}^d $ be a closed convex cone such that $\mathbb{R}^d_+\subseteq K$. The cone $K$ models the frictions between the markets. Specifically,  $K$ includes all positions that can be liquidated without any debt, thus is also known as solvency cone. For $1\leq p\leq\infty$, the set $L_d^p(K):=\{X\in L_d^p: X\in K, ~P\text{-}a.s.\}$ is a closed convex cone in $L_d^p$ generating a reflexive transitive relation for $\mathbb{R}^d$-valued random variables. 
	
	The closed convex cone $K$ also induces a partial ordering $\succeq$ on $\mathbb{R}^d$ by $x\succeq 0$ iff $x\in K$.  The partial ordering $\succeq$ can be extended naturally to $L_d^p$ by $X\succeq 0$ iff $X\in K$,  $P\text{-}a.s$.  The introduction of the partial order $\succeq$ makes it possible to compare more multivariate random variables.
	
	The value of a set-valued risk measure is the set of all deterministic portfolios that can compensate for the risk of the multivariate position. In practical applications, there are often certain requirements for the deterministic portfolios that can compensate, which are mainly characterized  through a set $M\subseteq \mathbb{R}^d$.  More precisely,  $M\subseteq \mathbb{R}^d$ is a linear subspace with dimension $1\leq m\leq d$. The introduction of $M$ means that an investor or regulator accepts the risk compensations or the security deposits only in a certain subset of the $d$ markets or currencies. For example,  only in the first $m$, in which case $M = \mathbb{R}^m \times \{0\}^{d-m}$. 
	
	The part of the cone $K$ that is relevant for $M$ is $K\cap M$, which is also a closed convex cone.  By int$(K\cap M)$ (or int$K \cap M$) we denote the interior of $K\cap M$ in $M$. Throughout the paper, it is assumed that $$\text{int}K\cap M\ne\emptyset,$$i.e., the interior of $K\cap M$ considered as a subset of the finite dimensional linear space $M$. Finally, we denote the collections of the upper closed subsets of $M$ by $$\mathbb{F}_M=\{D\subseteq M: D=\text{cl}(D+K\cap M)\}.$$ 
By the construction of $\mathbb{F}_M$, it is obviously $\emptyset \in \mathbb{F}_M$ and $K\cap M \in \mathbb{F}_M$. $K\cap M$  can be viewed as the building block for the element in $\mathbb{F}_M$. 
The multiplication is extended by $t\emptyset=\emptyset$ for $t>0$ and $0D=K \cap M$ for all $D\in \mathbb{F}_M$; in particular, $0\emptyset=K\cap M$, 
see more in \cite{2010dualityfor}.




	\subsection{Set-valued risk measures and its acceptance sets}
	
	Some basic definitions and facts of set-valued risk measures are recalled in this subsection.  The following definitions and key results about set-valued risk measures and acceptance sets are originated from \cite{2010dualityfor}. 
	
	More precisely, \textit{a set-valued risk measure} is a mapping $R : L_d^p\to \mathbb{F}_M$ which satisfies the following axioms:
	\begin{itemize}
		\item (R1) Cash Additivity : $R(X+u)=R(X)-u$ for any $X\in L_d^p$ and $u \in M$;
		\item (R2) $L_d^p(K)$-Monotonicity : $X-Y\in L_d^p(K)$ implies $R(X)\supseteq R(Y)$ for any $X, Y\in L_d^p$.
	\end{itemize}
	
	The value $R(X)$ of a set-valued risk measure $R$ encompasses the set of all eligible portfolio vectors that provide compensation for the risk associated with position $X$.  
	Besides,  a set-valued risk measure may also satisfy the following further properties as follows. 
	\begin{itemize}
		\item (R3) Normalization:  $K\cap M\subseteq R(0)$ and $R(0)\cap (- \mathrm{int}(K\cap M))=\emptyset$.
		
		\item (R4) Convexity : $tR(X)+(1-t)R(Y)\subseteq R(tX+(1-t)Y)$ for any $X, Y\in L_d^p$ and $t \in (0,1)$.
		
		\item (R5) Positive Homogeneity: $tR(X)=R(tX)$ for any $X\in L_d^p$ and $t>0$.
	\end{itemize}
	
	
	If a set-valued risk measure $R$ is normalized (respectively, convex, positive homogeneous), then it is called a (set-valued) normalized (respectively, convex, positive homogeneous) risk measure.  A coherent risk measure is a set-valued risk measure satisfying normalization, convexity and positive homogeneity. Since normalization only serves as a reference scale for the evaluation system and set-valued risk measures can be normalized through certain transformations, the definition of set-valued risk measures does not include the property of normalization in this paper.

	%

	Instead of considering set-valued risk measures directly, an alternative way of defining risk measures is provided by the notion of acceptance set, i.e., the set of random portfolios $X\in L_d^p$  which are viewed as free from risk by the supervisor or regulator.  
	
	We turn to properties of acceptance sets for set-valued risk measures.  \textit{An acceptance set} is a subset $A\subseteq L_d^p$ satisfying the following two basic properties:
	\begin{itemize}
		\item (A1) Closedness:  $A$ is directionally closed in $M$, i.e.,  $X\in L_d^p$, for all $\{u^k\}_{k\in \mathbb{N}}\subset M$ with $\lim_{k\rightarrow\infty}u^k=0$ and $X+u^k\in A$ for all $k\in\mathbb{N} $ implies $X\in A$.   Moreover,  $A+u\subseteq A$ whenever $u\in K\cap M$. 
		
		\item(A2) $L_d^p(K)$-Monotonicity: $A+L_d^p(K)\subseteq A$.
	\end{itemize}
	Besides, an acceptance set may satisfy the following further properties. 
	\begin{itemize}
		\item (A3) Normalization: $u\in K\cap M$ implies $u\in A$, and $u\in -(\mathrm{int}(K\cap M))$ implies $u\notin A$.
		
		\item (A4) Convexity: $tX+(1-t)Y\in A$ for all $X,Y\in A$ and all $t\in[0,1]$ .
		
		\item (A5) Cone: $tX\in A$ for all $t\in[0,+\infty)$ and all $X \in A$.
		
	\end{itemize}
	
	If an acceptance set $A$ is normalized (respectively, convex, normalized convex cone), then it is called a set-valued normalized (respectively, convex, coherent) acceptance set.
	
	It is known to all that the acceptance sets are intrinsically linked to the risk measures.  A set-valued risk measure provides the portfolios which compensate for the risk of a position, whereas a portfolio is an element of the acceptance set if its risk does not need to be compensated.  We associate with a mapping $R : L_d^p\to \mathbb{F}_M$ the set
	\begin{equation}\label{RtoA}
		A_R:=\{X\in L_d^p: 0\in R(X)\}.
	\end{equation}
	If $R$ is a set-valued risk measure, then $A_R$ includes those positions X which have zero among its risk compensating eligible portfolios, i.e., a position X is acceptable in terms of the risk measure $R$ if it can be made acceptable without additional initial endowment. 
	
	Conversely, let $A \subseteq L^p_d$ be a set.  Define
	\begin{equation}\label{AtoR}
		R_A(X):=\{u\in M: X+u\in A\},  \quad X\in L^p_d.
	\end{equation}
	Then $R_A$ maps $L_d^p$ into the power set of $M$.  If $A$ is an acceptance set, then the interpretation of (\ref{AtoR}) is, of course, that $R_A (X)$ includes all eligible portfolios which, when added to $X$, compensate for the risk of $X$. 
	
	As in the scalar case, \cite{2010dualityfor} showed that (\ref{RtoA}) and (\ref{AtoR}) yield  one-to-one correspondences between acceptance sets and set-valued risk measures.  
	
	\begin{lemma}\label{pro1}(\cite{2010dualityfor}, Proposition 3.1)
		(i) Let $R : L_d^p\to \mathbb{F}_M$ be a function satisfying (R1), then $A_R$ satisfies (A1) and $R=R_{A_R}$.
		
		(ii) Let $A\subseteq L_d^p$ be a set satisfying (A1), then $R_A$ maps into $\mathbb{F}_M$, satisfies (R1) and $A=A_{R_A}$.
	\end{lemma}
	\begin{lemma}\label{pro2}(\cite{2010dualityfor}, Proposition 3.2)
		(i) Let $R : L_d^p\to \mathbb{F}_M$ be a set-valued risk measure, then $A_R$ is an acceptance set. If $R$ is normalized (respectively, convex, positive homogenous), then $A_R$ is also normalized (respectively, convex, conical);
		
		(ii) Let $A\subseteq L_d^p$ be an acceptance set, then $R_A$ is a set-valued risk measure. If $A$ is normalized (respectively, convex, conical), then $R_A$ is also normalized (respectively, convex, positive homogenous).
	\end{lemma}
	The above two lemmas make it apparent which property of a (set-valued) translative function corresponds to what property of its zero sublevel set.
	
	\section{Set-valued star-shaped risk measures}\label{sec3}
	
	In this section, we will introduce the definition of set-valued star-shaped risk measures and set-valued star-shaped acceptance sets.  Moreover, the corresponding relationships between them are established.  
	
	\subsection{Star-shapedness}
	We first  give the definition of set-valued star-shaped risk measures and set-valued star-shaped acceptance sets in the following.   We focus on space $L_d^p$,  $1\leq p\leq\infty$, and the reader  can refer to \cite{2013timeconsistency} or \cite{2011setvaluedconical} for $L^{0}_{d}$ situation.
	
	\begin{definition}
		A set-valued star-shaped risk measure is a mapping $R : L_d^p\to \mathbb{F}_M$ satisfying (R1), (R2) and the following property:
		\begin{itemize}
			\item (R6) Star-shapedness : $tR(X)\subseteq R(tX)$ for all $X\in L_d^p$ and $t\in (0,1)$.
		\end{itemize}
		Besides, a star-shaped acceptance set is a subset $A\subseteq L_d^p$ satisfying (A1), (A2) and the following property:
		\begin{itemize}
			\item (A6) Star-shapedness: $tX\in A$ for all $X\in A$ and $t\in [0,1]$ .
		\end{itemize}
	\end{definition}
	
	The name ``star-shaped" is mainly expressed through the set of acceptable positions. (A6) means that for any point in a set of star-shaped, its connection to the origin is included in the set. Therefore,  a star-shaped set consists of a family of line segments with an origin as an endpoint, just like a star. The following lemma indicates that a set-valued star-shaped risk measure satisfies a half of normalization property. 
	
	\begin{lemma}\label{lem:3.1km}
		Let $R : L_d^p\to \mathbb{F}_M$ be a set-valued star-shaped risk measure, then we have that 
		$$0\in K\cap M\subseteq R(0).$$In particular,  (R6) is equivalent to the following (R6'):
		\begin{itemize}
			\item (R6') Star-shapedness : $tR(X)\subseteq R(tX)$ for all $X\in L_d^p$ and $t\in [0,1]$.
		\end{itemize}
	\end{lemma}	
	\begin{proof}
		Let $\{t_n\}_{n\in \mathbb{N}}\subset (0,1)$ be a sequence with $t_n\rightarrow 0$.  For any $X\in R(0)$, by the star-shapedness of $R$, 
		then it implies that 
		$$t_nX\in t_n R(0)\subseteq R(0).$$Since $t_nX\rightarrow 0$ as $t_{n}\rightarrow 0$, then the closeness of $R(0)$ means that  $0\in R(0)$.  Furthermore, upper closeness ensures that $0\in K\cap M\subseteq R(0)$.  
		
		The equivalence between (R6) and (R6') is from the fact that  $0D=K\cap M$ for any  $D\in \mathbb{F}_M$. \end{proof}

	\begin{remark}
		In the case of scalar-valued, $t$ can usually take the value $0$ in positive homogeneity. However, in (R5),  $R(0)=K\cap M$ when $t=0$, which is too strong requirement in the set-valued situation,  and may actually be taken as a closed convex cone. Similar to the result of Lemma
		\ref{lem:3.1km},  one can show that positive homogeneity (R5) of a set-valued star-shaped risk measure also implies that $K\cap M$ is included in $R(0)$.  In order to maintain consistency with the definitions of convexity and positive homogeneity,  $t$ in (R6) also takes in the values $(0,1)$. 
	\end{remark}
	
	\begin{remark}\label{remark2}
		
		Star-shapedness is a weakening of positive homogeneity and convexity in some sense.  Obviously,  positive homogeneity implies star-shapedness of a set-valued risk measure $R$.  Moreover,  convexity together with $0\in R(0)$ implies star-shapedness of $R$.  Indeed,  for any $X\in L_d^p$ and $t\in (0,1)$, 
		\begin{align*}
			tR(X)&=tR(X)+0\\
			&\subseteq tR(X)+(1-t)R(0)\\
			&\subseteq R(tX+(1-t)0)\\
			&=R(tX).
		\end{align*}
		Therefore, star-shapedness of a set-valued risk measure can be considered as convexity at zero point. The converse is generally not true, but it is true under the assumption of subadditivity,  see Proposition \ref{sub+xing}.
		
	\end{remark}

	\begin{remark}\label{yiwei}
		Let us specialize the discussion to the one-dimensional setting $d=m=1$. For any given set-valued star-shaped risk measure $R:L^p\to\mathbb{F}_M$, define
		$$\rho(X):=\inf R(X), ~~~\forall X\in L^p.$$
		Then for any  $X\in L^p$ and $t \in (0,1)$, we have that 
		$$\rho(tX)=\inf R(tX)\leq \inf tR(X)=t\inf R(X)=t\rho (X),$$
		i.e. the corresponding scalar-valued risk measure $\rho$ is  generalized star-shaped (it can take  the value $-\infty$).
	\end{remark}

We end this subsection by presenting some examples of  set-valued star-shaped risk measures. The readers can refer to \cite{2010dualityfor} or \cite{2021starshaped} for more examples and implications for set-valued risk measures or star-shaped risk measures. 	
	
	\begin{example}(Aggregations of convex risk measures)\label{exampleaggregation}
		\cite{2021starshaped} considered  the aggregation of scalar-valued star-shaped risk measures. In the set-valued case, consider the aggregation of set-valued convex risk measures $\{R_i:i\in I\}$: for any  $X\in L_d^p$, define
		$$R_{\min }(X)=\bigcap _{i\in I}R_i(X),$$
		$$R_{\max}(X)=\bigcup _{i\in I}R_i(X).$$
		$R_{\min }(X)$ is obviously convex, while $R_{\max}(X)$ satisfies star-shapedness but not convexity. Since both cases are too extreme, the convex combination of the two rules can be further defined:
		$$R(X)=\mu R_{\max}(X)+(1-\mu) R_{\min}(X),$$
		where the weight $\mu\in[0,1]$. In general, $R(\cdot)$ is a star-shaped but not convex risk measure.
		
		On the other hand, it can be easily verified that for a collection of set-valued star-shaped risk measures, their convex combination, union and intersection are also star-shaped risk measures.
	\end{example}
	
 \begin{example} (Set-valued $V@R$, \cite{2010dualityfor})
        Let $0\leq \lambda \leq 1$, and $\mathcal{Q}$ be a family of probability measures. For any $Q\in\mathcal{Q} $, mapping $X\mapsto V@R_\lambda^Q$ is defined as 
		$$V@R_\lambda^Q(X)=\{u\in M:Q(X+u\notin K)\leq\lambda\}, ~~~X\in L_d^p.$$
		$V@R_\lambda^Q$ is called set-valued value at risk. 
 It is a set-valued normalized risk measure satisfying positive homogeneity, thus is star-shaped, but not convex in general. 
 
 Based on this, similar to  \cite{wang2021expected}, we can consider scenario robust set-valued $V@R$. Specifically, $$MaxV@R_\lambda^\mathcal{Q}(X)=\bigcup_{Q\in\mathcal{Q}}V@R_\lambda^Q(X),~~~~X\in L_d^p.$$It is usually not convex, but is positive homogeneity and star-shaped.\end{example}

\begin{example}(Nonconcave Utilities) Motivated by \cite{2021starshaped}, we can define set-valued shortfall risk measure based on nonconcave utility:
    \begin{align}\label{shortfall}
    R_u(X)=\left\{v\in M:\mathbb{E}[u(-X-v)]\leq u(0)\right\}, ~~\forall X\in L_d^p,
    \end{align}
    where $u$ is a suitable increasing and nonconstant utility function on $\mathbb{R}$ such that $u(0)=0$ Under risk measure $R_u(\cdot)$, the acceptable risk positions are the ones that have nonnegative reservation price. Similar to \cite{2016Stochastic}, risk measures defined by (\ref{shortfall}) is convex if the utility function $u$ is concave. Consider a kind of special nonconcave utility satisfying
    \begin{align}\label{utility}
        \lambda\mapsto\frac{u(\lambda x)}{\lambda} \text{~~is  decreasing on~~} (0,\infty) ~\text{~~for~~}  x\in\mathbb{R}.
    \end{align}
    If $u$ satisfies \eqref{utility}, then $R_u$ is star-shaped. For $\lambda\in(0,1)$, $X\in L_d^p$,
    \begin{align*}
        R_u(\lambda X)&=\{\lambda v\in M:\mathbb{E}[u(-\lambda X-\lambda v)]\leq 0\}\\
        &=\lambda\left\{ v\in M:\mathbb{E}[\frac{u(-\lambda X-\lambda v)}{\lambda}]\leq 0\right\}\\
        &\subseteq\lambda\{ v\in M:\mathbb{E}[u(-X-v)]\leq 0\}\\
        &=\lambda R_u(X).
    \end{align*}
    In fact, it is not difficult to verify that the risk measure defined by \eqref{shortfall} is a set-valued star-shaped risk measure if $u$ satisfies \eqref{utility}.
    \end{example}

	\subsection{Star-shaped risk measures and acceptance sets}	
	In this subsection,  we present some properties of star-shaped risk measures and their relationship with the star-shaped acceptance sets.

	The following proposition establishes the connection of star-shaped risk measure with risk-to-exposure ratio.
	
	\begin{proposition}\label{pro: 3.2}
		For a set-valued risk measure $R : L_d^p\to \mathbb{F}_M$, the following are equivalent:
		
		(i) $R$ is star-shaped;
		
		(ii) $R(tX)\subseteq tR(X)$ for any $X\in L_d^p$ and $t\in (1,\infty)$;
		
		(iii) For each $X\in L_d^p$, the risk-to-exposure ratio $r_x: \beta \to R(\beta X)/\beta$ is a shrinking mapping of $\beta$ on $(0,\infty)$, i.e. $\beta_1>\beta_2>0$ implies $R(\beta_1 X)/\beta_1\subseteq R(\beta_2 X)/\beta_2.$ 
		
	\end{proposition}
	
	\begin{proof}
		(i) implies (ii).  If $t\geq 1$,  then $1/t \in (0,1)$. For all $X\in L_d^p$, star-shapedness then implies
		$$R(X)=R(\frac{1}{t}(tX))\supseteq\frac{1}{t}R(tX).$$
		
		(ii) implies (iii). Let $\beta_1>\beta_2>0$, then $\beta_1/\beta_2>1$. For any $X\in L_d^p$, one gets that
		$$R(\beta_1 X)=R(\frac{\beta_1}{\beta_2}(\beta_2 X))\subseteq\frac{\beta_1}{\beta_2}R(\beta_2 X),$$
		that is $R(\beta_1 X)/\beta_1\subseteq R(\beta_2 X)/\beta_2.$
		
		(iii) implies (i). For all $X\in L_d^p$ and all $\lambda\in (0,1)$, because $r_X$ is reducing, one has
		$$R(X)=\frac{R(X)}{1}\subseteq \frac{R(\lambda X)}{\lambda}$$as wanted.
	\end{proof}

	According to Proposition \ref{pro: 3.2} (iii), we can see that the rate of increase in star-shaped risk measure is slower as the position increases, which is due to the presence of liquidity risk.  Star-shapedness reflects the concentration of assets and liquidity issues.
	
	For acceptance sets, there are also corresponding equivalent forms. We omit its proof.
	\begin{proposition}
		For an acceptance set $A\subseteq L_d^p$, the following are equivalent:
		
		(i) $A$ is star-shaped;
		
		(ii) $tA\subseteq A$ for all $t\in [0,1]$;
		
		(iii) $A\subseteq tA$ for all $t\in [1,\infty)$.
	\end{proposition}

	Similar to scale-valued case, the following proposition presents an equivalent relationship between star-shapedness, positive homogeneity and convexity in the set-valued situation.
	
	\begin{proposition}\label{sub+xing}
		For a normalized subadditive risk measure $R: L_d^p\to \mathbb{F}_M$, where subadditivity means that  $R(X)+R(Y)\subseteq R(X+Y)$ for any $X,  Y\in L_d^p$.  Then  the following are equivalent:
		
		(i) $R$ is star-shaped;
		
		(ii) $R$ is positive homogeneous;
		
		(iii) $R$ is convex.
	\end{proposition}
	\begin{proof}
		(i) implies (ii). Let $X\in L_d^p$. On the one hand, subadditivity of $R$ implies that
		$$R(2^nX)=R(2^{n-1}X+2^{n-1}X)\supseteq 2R(2^{n-1}X)$$
		for all $n\in \mathbb{Z}$. On the other hand,  by Proposition \ref{pro: 3.2}, $r_x: \beta \to R(\beta X)/\beta$ is a shrinking mapping of $\beta$ on $(0,\infty)$, then one can get that 
		$$r_x(2^n)=\frac{R(2^n X)}{2^n}\supseteq \frac{R(2^{n-1} X)}{2^{n-1}}=r_x(2^{n-1})\supseteq r_x(2^n).$$
		
		Therefore, the shrinking mapping $r_x$ is an identity mapping on the set $\{2^n:n\in\mathbb{Z}\}$.  This makes 
		$r_x(\lambda)=R(X)$ for all $\lambda>0$, which means it is an invariant set on $(0,\infty)$ and it implies positive homogeneity.
		
		The rest claims are straightforward verified by Remark \ref{remark2}.
	\end{proof}
	
	A one-to-one correspondence between set-valued risk measures and acceptance sets is established in Lemma \ref{pro2}.  This one-to-one correspondence also holds for star-shapedness.
	
	\begin{proposition}\label{pro3}
		(i) Let $R : L_d^p\to \mathbb{F}_M$ be a set-valued star-shaped risk measure, then $A_R$ is a star-shaped acceptance set.
		
		(ii) Let $A\subseteq L_d^p$ be a star-shaped acceptance set, then $R_A$ is a set-valued star-shaped risk measure. 
	\end{proposition}
	\begin{proof}
		From Lemma \ref{pro2}, we only need to show that star-shapedness of $R$ implies star-shapedness of $A_R$ and star-shapedness of $A$ implies star-shapedness of $R_A$.
		
		$(i).$  Let $R : L_d^p\to \mathbb{F}_M$ be a set-valued star-shaped risk measure. Take $X\in A_R$ and $t\in(0,1)$. The construction (\ref{RtoA}) implies $0\in R(X)$. Then $$0=t0\in tR(X)\subseteq R(tX),$$ and hence $tX\in A_R$ for $t\in(0,1)$ as desired. $1X\in A_R$ is apparent and $0\in A_R$ is immediate from Lemma \ref{lem:3.1km}.
		
		$(ii).$   Let $A\subseteq L_d^p$ be a star-shaped acceptance set. Take $X\in A_R$ and $t\in(0,1)$.
		\begin{align*}
			tR_A(X)&=t\{u\in M:X+u\in A\}\\
			&=\{tu\in M:X+u\in A\}\\
			&\subseteq \{tu\in M:tX+tu\in A\}\\
			&=\{v\in M:tX+v\in A\}\\
			&=R_A(tX)	 	
		\end{align*}
		where the inclusion holds true since $X+u\in A$ implies $tX+tu\in A$ by (A6).
	\end{proof}

    \section{Representations of set-valued risk measures and star-shaped normalized risk measures}\label{sec4}
	
	In the scalar-valued case,   
    \cite{2020monetary} showed that a scalar-valued monetary risk measure is the lower envelope of a family of convex risk measures. \cite{2021starshaped} provided that a scalar-valued star-shaped normalized risk measures is the minimum of the collection of all convex risk measures that dominate it. 
	
	Motivated by the results of scalar-valued case, this section provides the representation theorems for set-valued situations. Specifically, in the field of set-valued risk measures, these risk measures can still be represented in terms of a family of set-valued convex risk measures, but no longer by their lower envelope or minimum, but by their union.
	
	Due to the one-to-one correspondence between risk measures and acceptable sets, the research on risk measures can be transformed into the study of acceptable sets. By splitting the acceptable set of the target risk measures into a family of convex (convex normalized) acceptable sets,  then the corresponding relationship between risk measures can be established.  It should be noted that,  different from the ``scalar-valued'' case,  the ``set-valued '' characterizations of the risk measures brings some new difficulties and challenges in the construction of the acceptable sets.    
	
	\subsection{Representations of the set-valued risk measures}
	
	The following theorem shows that a set-valued risk measure can be represented by a family of set-valued convex risk measures.
	
	\begin{theorem}\label{theo2}
		For a given mapping $R : L_d^p\to \mathbb{F}_M$, the following statements are equivalent.
		
		(i)  $R$ is a set-valued risk measure.
		
		(ii)  There exists a family of convex acceptance sets $\{A_\gamma:\gamma\in \Gamma\}$ such that
		$$R(X)=\{u\in M: X+u\in A_\gamma\ {\rm for\ some}\ \gamma\in \Gamma\},~~ \forall X\in L_d^p.$$
		
		(iii)  There exists a family of set-valued convex risk measures $\{R_\lambda:\lambda \in\Lambda\}$  on $L_d^p$ such that
		$$R(X)=\bigcup_{\lambda \in\Lambda}R_\lambda(X), \forall X\in L_d^p.$$
		
		(iv) For any $X\in L_d^p,$
		$$R(X)=\bigcup \{R_\xi(X)~| ~R_\xi\text{ is a  set-valued convex  risk  measure and }  R_\xi \subseteq R\}.$$		
		
	\end{theorem}
	
	\begin{proof}
		(i) implies (ii).  Let $R$ be a set-valued risk measure and $A_R$ be the acceptance set of $R$.   For any $Z\in A_R$,  let
		\begin{align}
			A(Z):=\{X\in L_d^p: X\succeq Z\}=\{Z\}+L_d^p(K).
		\end{align}
		Then each $A(Z)$ is a convex subset of $L_d^p$ satisfying (A1) and (A2), i.e. $A(Z)$ is a convex acceptance set. Since $A_R$ satisfies (A2),  then we have 
		$$A(Z)=\{Z\}+L_d^p(K)\subseteq A_R, ~~~\forall Z\in A_R.$$ Therefore, it is obvious that $$A_R=\bigcup_{Z\in A_R}\{ Z\}\subseteq\bigcup_{Z\in A_R}A(Z)\subseteq A_R,$$which means that
		$$A_R=\bigcup_{Z\in A_R}A(Z).$$ 
		
		Hence, for any $X\in L_d^p$,  Lemma \ref{pro1} implies that
		\begin{align*}
			R(X)=R_{A_R}(X)&=\{u\in M: X+u\in A_R\}\\
			&=\{u\in M: X+u\in \bigcup_{Z\in A_R}A(Z)\}\\
			&=\{u\in M: X+u\in A(Z) {\rm\ for\ some\ }Z\in A_R\}.  
		\end{align*}
		Then, $\{A(Z):Z\in A_R\}$ is the desired family of convex acceptance sets.
		
		(ii) implies (iii). Let $\{A_\gamma:\gamma\in \Gamma\}$ be a family of convex acceptance sets such that
		$$R(X)=\{u\in M: X+u\in A_\gamma\mathrm{\ for\ some\ } \gamma\in \Gamma\}, ~~\forall X\in L_d^p.$$
		Denote by $R_\gamma$ the correspond set-valued risk measure of each $A_\gamma$ for $\gamma\in\Gamma$. By Lemma \ref{pro2}, then each $R_\gamma$ is a set-valued convex risk measure.
		
		For all $X\in L_d^p$,  one  can get that
		\begin{align*}
			R(X)&=\{u\in M: X+u\in A_\gamma \mathrm{\ for\ some\ } \gamma\in \Gamma\}\\
			&=\{u\in M: X+u\in \bigcup_{\gamma\in \Gamma}A_\gamma \}\\
			&=\bigcup_{\gamma\in \Gamma}\{u\in M: X+u\in A_\gamma\}\\
			&=\bigcup_{\gamma\in \Gamma}R_\gamma(X).
		\end{align*}
		Thus, $\{R_\gamma:\gamma\in \Gamma\}$ is the desired family of set-valued convex risk measures. 
		
		(iii) implies (iv). Let $\{R_\lambda:\lambda \in\Lambda\}$ be the family of set-valued convex risk measures on $L_d^p$ such that
		$$R(X)=\bigcup_{\lambda \in\Lambda}R_\lambda(X),~~ \forall X\in L_d^p.$$
		Let $\{R_\xi:\xi \in\Xi\}$ be the all set-valued convex risk measures  on $L_d^p$ that dominated by $R$, that is to say, $$R_\xi(X)\subseteq R(X), ~\forall \xi\in\Xi,  ~\forall X\in L_d^p.$$ Since $R(X)$ is closed, then $$R(X)=\bigcup_{\lambda \in\Lambda}R_\lambda(X)\subseteq\bigcup_{\xi\in \Xi}R_{\xi}(X)\subseteq R(X), ~~\forall X\in L_d^p,$$ which is equivalent to that,  for any $X\in L_d^p$,
		\begin{align*}
			R(X)&=\bigcup_{\xi\in \Xi}R_{\xi}(X)\\
			&=\bigcup \{R_\xi(X)~| ~R_\xi \text{ ~is a  set-valued convex risk  measure and ~}  R_\xi \subseteq R\}.
		\end{align*}
		
		(iv) implies (i). Let $\{R_\xi:\xi \in\Xi\}$ be the all set-valued convex measures of risk on $L_d^p$ that dominated by $R$ such that
		$$R(X)=\bigcup_{\xi \in\Xi}R_\xi(X), ~~\forall X\in L_d^p.$$
		We need to prove that $R$ satisfies (R1) and (R2).
		
		To check (R1), take $X\in L_d^p$ and $u \in M: $
		$$R(X+u)=\bigcup_{\xi \in\Xi}R_\xi(X+u)=\bigcup_{\xi \in\Xi}(R_\xi(X)-u)=\bigcup_{\xi \in\Xi}R_\xi(X)-u=R(X)-u.$$
		For (R2), taking $X$ and $Y\in L_d^p$ such that $Y-X\in L_d^p(K)$,
		$$R(Y)=\bigcup_{\xi \in\Xi}R_\xi(Y)\supseteq\bigcup_{\xi \in\Xi}R_\xi(X)=R(X).$$
	\end{proof}
	\begin{remark}
It is noting that in the scalar situation, \cite{2013riskpreferences}  provided the representations for general quasi-convex risk measures by risk acceptance set family. Part (ii) of Theorem \ref{theo2} presented  the similar representation for the set-valued risk measures.  

	\end{remark}
	
	\begin{remark}
		$A(Z)$ is the smallest acceptance set containing $Z$.  The corresponding set-valued risk measure is 
		\begin{align*}
			R_{A(Z)}(X)&=\{u\in M: X+u\in A(Z)\}\\
			&=\{u\in M: X+u\in \{Z\}+L_d^p(K)\}\\
			&=\{u\in M: X-Z+u\in L_d^p(K)\}\\
			&=WC^{M,S}(X-Z),
		\end{align*}
		where $WC^{M,S}$ is called set-valued worst case risk measure. Together with Theorem 6.1 in \cite{2010dualityfor},  we can obtain the following necessary condition for set-valued risk measures.
	\end{remark}
	
	\begin{corollary}
		A set-valued risk measure $R: L_d^p\to \mathbb{F}_M$ has the following representation	
		$$R(X)=\bigcup_{Z\in A_R}\bigcap_{(Q, y)\in\mathcal{W}}\big(E^Q([-(X-Z)])+G(y)\big)\cap M, ~~\forall X\in L_d^p,$$
		where $$\mathcal{W}=\big\{(Q, y)\in\mathcal{M}_{1,d}^P\times K^+\backslash M^\bot~:~{\rm diag}(y)\frac{dQ}{dP}\in L_d^p(K^+)\big\}.$$
		$\mathcal{M}_{1,d}^P$ is the set of all vector probability measures with components absolutely continuous with respect to $P$. Here,  $K^+$ is the positive dual cone of the cone $K$ in $\mathbb{R}^d$, $M^\bot=\{v\in\mathbb{R}^d:\forall u\in M: v^Tu=0\}$ and $G(y)=\{x\in \mathbb{R}^d: y^Tx\geq 0\}$.	
	\end{corollary}
	
	\begin{remark}
		Similar as Remark \ref{yiwei}, let us specialize the Theorem \ref{theo2} to the one-dimensional setting $d=m=1$. For any given set-valued risk measure $R:L^p\to\mathbb{F}_M$, define
		$$\rho(X):=\inf R(X), ~~~\forall X\in L^p.$$ Then $\rho$ is a monetary risk measure, and by Theorem \ref{theo2}(iii), there exists a family of set-valued convex risk measures $\{R_\lambda:\lambda \in\Lambda\}$  on $L_d^p$ such that
		$$R(X)=\bigcup_{\lambda \in\Lambda}R_\lambda(X),  ~~\forall X\in L_d^p.$$
		Thus,  for any  $X\in L^p$, we have that 
		$$\rho(X):=\inf R(X)=\inf \bigcup_{\lambda \in\Lambda}R_\lambda(X)=\min(\inf_{\lambda\in\Lambda}R_\lambda(X))=\min\rho_\lambda(X) $$
		where $\rho_\lambda(X):=\inf_{\lambda\in\Lambda}R_\lambda(X), \lambda\in \Lambda$ is the desired family of scalar-valued convex risk measures and is correspond with the representation theorem in \cite{2020monetary}.
	\end{remark}
	
	\subsection{Representations of the set-valued star-shaped normalized risk measures}
	The following theorem shows that a set-valued star-shaped normalized risk measure can be represented by a family of set-valued convex normalized risk measures.
	\begin{theorem}\label{theo1}
		For a mapping $R : L_d^p\to \mathbb{F}_M$, the following statements are equivalent:
		
		(i)  $R$ is a set-valued star-shaped normalized risk measure.
		
		(ii)  There exists a family of convex normalized acceptance sets $\{A_\gamma:\gamma\in \Gamma\}$ such that
		$$R(X)=\{u\in M: X+u\in A_\gamma{\rm\ for\ some}\ \gamma\in \Gamma\}, ~~\forall X\in L_d^p.$$
		
		(iii)  There exists a family of  set-valued convex normalized risk measures $\{R_\lambda:\lambda \in\Lambda\}$ on $L_d^p$ such that
		$$R(X)=\bigcup_{\lambda \in\Lambda}R_\lambda(X), ~~\forall X\in L_d^p.$$
		
		(iv) For any $X\in L_d^p,$
		$$R(X)=\bigcup \{R_\xi(X)~|~ R_\xi  \text{ is  a  set-valued  convex normalized risk measure and }  R_\xi \subseteq R\}.$$		
		
	\end{theorem}
	\begin{proof}
		We only prove ``(i)$\Rightarrow$(ii)" and ``(iv)$\Rightarrow$(i)", the remainder of the arguments  can be verified analogous to that in Theorem \ref{theo2}.
		
		(i) implies (ii).  Let $R$ be a star-shaped normalized risk measure, and let $A_R$ be the acceptance set of $R$, then $A_R$ is a star-shaped and normalized acceptance set.   
		
		For any $Z\in A_R$,  define
		\begin{align}
			A(Z):=\text{conv}(\{0,Z\}+L_d^p(K))=\bigcup_{t\in [0,1]}(\{tZ\}+L_d^p(K))=\bigcup_{t\in [0,1]}\{tZ\}+L_d^p(K),
		\end{align}where $\text{conv}(\cdot)$ stands for a convex hull of a set.  	
		Then each $A(Z)$ is a convex subset of $L_d^p$ satisfying (A1) and (A2) and (A3), i.e. $A(Z)$ is a convex normalized acceptance set.  By (A2) and convexity of $A_R$, we have 
		$$A(Z)=\bigcup_{t\in [0,1]}\{tZ\}+L_d^p(K)\subseteq A_R, ~~\forall Z\in A_R.$$ On the other hand, we also have  $$A_R=\bigcup_{Z\in A_R}\{ Z\}\subseteq\bigcup_{Z\in A_R}A(Z)\subseteq A_R,$$and it is equivalent to 
		$$A_R=\bigcup_{Z\in A_R}A(Z).$$ 
		
		Therefore, for any $X\in L_d^p$,  Lemma \ref{pro1} implies that
		\begin{align*}
			R(X)=R_{A_R}(X)&=\{u\in M: X+u\in A_R\}\\
			&=\{u\in M: X+u\in \bigcup_{Z\in A_R}A(Z)\}\\
			&=\{u\in M: X+u\in A(Z) {\rm\ for\ some\ }Z\in A_R\},
		\end{align*}
		where  $\{A(Z):Z\in A_R\}$ is the desired family of convex normalized acceptance sets.
		
		(iv) implies (i). Let $\{R_\xi:\xi \in\Xi\}$ be all set-valued convex normalized measures of risk on $L_d^p$ that dominated by $R$ such that
		$$R(X)=\bigcup_{\xi \in\Xi}R_\xi(X), ~~\forall X\in L_d^p.$$
		We need to prove that $R$ satisfies (R1), (R2), (R3) and (R6). 
		
		The properties (R1) and (R2) can be easily verified similar to the situation in Theorem 
		\ref{theo2}.
				
		To check (R3), choose $ \xi \in \Xi$, since $K\cap M\subseteq R_\xi(0)$ and $R_\xi(0)\cap - {\rm int}\ (K\cap M)=\emptyset$, then $K\cap M\subseteq \bigcup_{\xi \in\Xi}R_\xi(0)=R(0)$ and $$R(0)=\bigcup_{\xi \in\Xi}R_\xi(0)\subseteq (- {\rm int}\ (K\cap M))^c,$$
		which implies that $R(0)\cap - {\rm int}\ (K\cap M)=\emptyset$.
		
		To check (R6), take $X\in L_d^p$ and $t\in (0,1),$  one gets that
		\begin{align*}
			tR(X)=t\bigcup_{\xi \in\Xi}R_\xi(X)&=\bigcup_{\xi \in\Xi}(tR_\xi(X)+0)\\
			&\subseteq\bigcup_{\xi \in\Xi}(tR_\xi(X)+(1-t)R_\xi(0))\\
			&\subseteq\bigcup_{\xi \in\Xi}(R_\xi(tX+(1-t)0)\\
			&=\bigcup_{\xi \in\Xi}R_\xi(tX)=R(tX).
		\end{align*}
	\end{proof}
	
	\begin{remark}
		$A(Z)$ is also the smallest convex normalized acceptance set containing $Z$.  If $Z\in L_d^p(K)$, then $A(Z)=L_d^p(K)$. If $Z\notin L_d^p(K)$, then $A(Z)$ represents the risk positions that is lager than $tZ$ for some $t\in[0,1]$ under the partial ordering $\succeq$.
	\end{remark}
	
	In Example \ref{exampleaggregation}, union, intersection and convex combination of a family of set-valued star-shaped risk measures are also star-shaped risk measures. The following result states that Theorem \ref{theo1} performs well under aggregation, and its proofs are omit.
	\begin{proposition}
        Let $\{R_i\}_{i\in I}$ be a family of set-valued star-shaped risk measure.  For any $i\in I$, let $\Tilde{\mathcal{R}}_i$ (determined by Theorem \ref{theo1} (iv)) be the set of all set-valued convex normalization risk measures controlled by $R_i$. 
        Then, 
        union, intersection and convex combination of $\{R_i\}_{i\in I}$ are given respectively by
        $$\bigcap_{i\in I}R_i(X)=\bigcup_{\Tilde{R}\in\cap_i\Tilde{\mathcal{R}}_i}\Tilde{R}(X),$$
        $$\bigcup_{i\in I}R_i(X)=\bigcup_{\Tilde{R}\in\cup_i\Tilde{\mathcal{R}}_i}\Tilde{R}(X),$$
        $$\sum_{i\in I}t_iR_i(X)=\bigcup\left\{\Tilde{R}(X):\Tilde{R}\text{~is set-valued convex normalized risk measure and~}\Tilde{R}\in\sum_{i\in I}t_i\Tilde{\mathcal{R}}_i\right\},$$
        where $t_i\geq0,\sum_{i\in I} t_i=1.$
    \end{proposition}

	The following corollary shows that a positive homogeneous set-valued risk measure can be represented by a family of set-valued coherent risk measures.
	
	\begin{corollary}\label{coro1}
		For a mapping $R : L_d^p\to \mathbb{F}_M$, the following statements are equivalent:
		
		(i)  $R$ is a positive homogeneous set-valued risk measure.
		
		(ii)  There exists a family of coherent acceptance sets $\{A_\gamma:\gamma\in \Gamma\}$ such that
		$$R(X)=\{u\in M: X+u\in A_\gamma\ {\rm for\ some}\ \gamma\in \Gamma\}, \forall X\in L_d^p.$$
		
		(iii)  There exists a family of set-valued coherent risk measures $\{R_\lambda:\lambda \in\Lambda\}$ on $L_d^p$ such that
		$$R(X)=\bigcup_{\lambda \in\Lambda}R_\lambda(X), \forall X\in L_d^p.$$
		
		(iv) ~ $\forall X\in L_d^p,$ we have
		$$R(X)=\bigcup \{R_\xi(X)~| ~R_\xi  \text{ is  a set-valued  coherent  risk  measure and }  R_\xi \subseteq R\}.$$		
		
	\end{corollary}
	\begin{proof}
		We only prove ``(i)$\Rightarrow$(ii)",  the remainder of the arguments  can be verified analogous to that in Theorem \ref{theo2}.
		
		(i) implies (ii). Let $R$ be a positive homogeneous set-valued risk measure, and let $A_R$ be the acceptance set of $R$, then $A_R$ is a cone. 
		
		For any $Z\in A_R$, define
		\begin{align}
			A(Z):=\bigcup_{t\in [0,\infty)}\{tZ\}+L_d^p(K).
		\end{align}
		Then each $A(Z)$ is a convex cone of $L_d^p$ satisfying (A1) and (A2) and (A3), i.e. $A(Z)$ is a coherent acceptance set.   Hence, by (A2) and positive homogeneity of $A_R$,  we have 
		$$A(Z)=\bigcup_{t\in [0,\infty)}\{tZ\}+L_d^p(K)\subseteq A_R, ~~\forall Z\in A_R,$$and furthermore,  
		
		$$A_R=\bigcup_{Z\in A_R}A(Z).$$ 
		
		Therefore, for any $X\in L_d^p$,  Lemma \ref{pro1} implies that
		\begin{align*}
			R(X)=R_{A_R}(X)&=\{u\in M: X+u\in A_R\}\\
			&=\{u\in M: X+u\in \bigcup_{Z\in A_R}A(Z)\}\\
			&=\{u\in M: X+u\in A(Z) {\rm\ for\ some\ }Z\in A_R\},
		\end{align*}
		where $\{A(Z): Z\in A_R\}$ is the desired family of coherent acceptance sets. \end{proof}

	\begin{example} (Set-valued value at risk, \cite{2010dualityfor})  Let $0\leq \lambda \leq 1$. The functions $X\mapsto V@R_\lambda^{M,W}$ and  $V@R_\lambda^{M,S}:L_d^p\to \mathbb{F}_M$ are respectively defined by
		$$V@R_\lambda^{M,W}(X)=\{u\in M:P(X+u\in-{\rm int} K)\leq\lambda\}$$
		and
		$$V@R_\lambda^{M,S}(X)=\{u\in M:P(X+u\notin K)\leq\lambda\}.$$
		$V@R_\lambda^{M,W}$ and $V@R_\lambda^{M,S}$ are set-valued normalized risk measures, while they cannot be convex in general, but they do satisfy positive homogeneity.   By Theorem \ref{coro1}, we have
		$$V@R_\lambda^{M,W}(X)=\bigcup \{R(X)~|~ R \text{ is  a set-valued  coherent  risk  measure and }   R\subseteq V@R_\lambda^{M,W}\}$$		
		and 
		$$V@R_\lambda^{M,S}(X)=\bigcup \{R(X)~|~ R \text{ is  a set-valued  coherent  risk  measure and }  R\subseteq V@R_\lambda^{M,S}\}.$$		
		The above results are consistent with the case of scalar-valued case that
		$$VaR_\alpha(X) = \inf \{h(X)~ | ~h \text{ is  a scalar-valued  coherent  risk  measure and }   h \geq VaR_\alpha\}.$$
		
	\end{example}

	\begin{remark}
		By Remark \ref{remark2} and Lemma \ref{lem:3.1km}, we know that under the assumption of convexity,  $0\in R(0)$ is equivalent to star-shapedness. In fact, instead of demanding for each $R_\lambda$ to be normalized in Theorem \ref{theo1}, we could simply ask them to satisfy $0\in R_{A_\lambda}(0)$.
	\end{remark}
	
	The following proposition provides a necessary condition for star shapedness of set-valued risk measures.
	
	\begin{proposition}\label{pro4.1}
		Let $R$ be a set-valued risk measure. If $R$ is star-shaped, then there exists a family  of set-valued convex risk measures $\{R_\lambda:\lambda\in\Lambda\}$ with at least one member that is star-shaped such that
		$$R(X)=\bigcup_{\lambda \in\Lambda}R_\lambda(X), ~~\forall X\in L_d^p.$$
	\end{proposition}
	
	\begin{proof}
		By Theorem \ref{theo2},  there exists a family of set-valued convex risk measures $\{R_\lambda:\lambda\in\Lambda\}$ such that
		$$R(X)=\bigcup_{\lambda \in\Lambda}R_\lambda(X), ~~\forall X\in L_d^p.$$   We only need to show that $\{R_\lambda:\lambda\in\Lambda\}$ contains a star-shaped risk measure. By Remark \ref{remark2}, it suffices to  show  that there exists $\lambda \in\Lambda$ such that $0\in R_\lambda (0)$. As $R$ is star-shaped, $0\in R(0)=\bigcup_{\lambda \in\Lambda}R_\lambda(0)$. Thus,  $0\in R_{\lambda^*}(0)$ for some $\lambda^* \in\Lambda$ and $R_{\lambda^*}$ is star-shaped.
	\end{proof}	
	
	\begin{remark}
	    The converse of Proposition \ref{pro4.1} no longer holds generally. Such representation could only guarantee that $R$ is a set-valued risk measure satisfying $0\in R(0)$. If we require $0\in R_\lambda (0)$ for all $\lambda\in\Lambda$, then one can get $R$ is star-shaped.
	\end{remark}
	
	\section{On the link between set-valued risk measures and star-shaped risk measures}\label{sec5}
	
	\cite{2022montaryandstarshaped} found a subtle relationship between the results of \cite{2020monetary} and \cite{2021starshaped}.  In this section, we provide a similar set-valued version about the relationship between set-valued risk measures and set-valued star-shaped risk measures.  
	
	In order to study the link between set-valued risk measures and star-shaped risk measures, one first needs to introduce a broader definition of star-shapedness.
	
	\begin{definition}
		A nonempty set $A\subseteq L_d^p$ is star-shaped at $B\subseteq L_d^p$ if $tX+(1-t)b\in A$ for any  $X\in A$, $b\in B$ and $t\in  [0,1]$. Particularly,  if $A$ is star-shaped at a singleton $\{Y\}$,  we say that $A$ is star-shaped at $Y$. Unless otherwise specified,  star-shapedness is to be understood as star-shapedness at $0$.
	\end{definition}
	
	\begin{remark}
		The definition of star-shapedness implies that if $A$ is star-shaped at $B\subseteq L_d^p$, then $B$ is a subset of $A$.
	\end{remark}
	
    \begin{definition}\label{def5.2}For some given $Y\in L_d^p$ and a risk measure $R$ with $0\in R(Y)$, we say $R$ is star-shaped at $Y$,  if $$tR(X)+(1-t)R(Y)\subset R(tX+(1-t)Y), ~~\forall X\in L_d^p,  ~t\in[0,1].$$
\end{definition}
	
The following proposition provides a more elaborate representation for the set-valued risk measure with its acceptance set star-shaped at some point.

	\begin{proposition}\label{pro5.2}
		Let $R$ be a set-valued risk measure,  and $R$ is star-shaped at $Y\in L_d^p$. Then there exists a family  of convex acceptance sets $\{A_\lambda:\lambda\in\Lambda\}$ such that $A_R=\bigcup_{\lambda\in \Lambda} A_\lambda$, $\bigcap_{\lambda\in \Lambda} A_\lambda\neq \emptyset$ and 
		$$R(X)=\bigcup_{\lambda \in\Lambda}R_{A_\lambda}(X), ~~\forall X\in L_d^p.$$
	\end{proposition}
	\begin{proof}
		By Theorem \ref{theo2}, we know that there exists a family $\{A'_\lambda:\lambda\in\Lambda\}$ of convex acceptance sets such that $A_R=\bigcup_{\lambda\in \Lambda} A'_\lambda$ and 
		$R(X)=\bigcup_{\lambda \in\Lambda}R_{A'_\lambda}(X)$, $\forall X\in L_d^p$.  However, we can not  guarantee $\bigcap_{\lambda\in \Lambda} A'_\lambda\neq \emptyset$.  
		
		Therefore, we need further construction and  to show that there exists a family of such sets with a non-empty intersection.    For each $Z\in L_d^p$,  let $$A(Z):=\{X\in L_d^p:X\succeq Z\}.$$Note that each of such sets is a convex acceptance set. 
  
 Since $R$ is star-shaped at $Y$, we can get $A_R$ is star-shaped at $Y$. Indeed, for any $X\in A_R$ and $t\in[0,1]$, by Definition \ref{def5.2}, $0\in tR(X)+(1-t)R(Y)\subset R(tX+(1-t)Y)$, i.e. $tX+(1-t)Y\in A_R$.
  Besides, $A_R=\bigcup_{Z\in A_R}A(Z)$. As $A_R$ is star-shaped at $Y$ and monotone, then it is also star-shaped at $A(Y)$.  Indeed,  for all $t\in[0,1]$, $X\in A_R$ and  $W\in A(Y)$, we have that  $W-Y\in L_d^p(K)$ and 
		\begin{align*}
			tX+(1-t)W&=tX+(1-t)Y+(1-t)(W-Y)\\
			&\in A_R+(1-t)L_d^p(K)\\
			&\subseteq A_R+L_d^p(K)\\
			&\subseteq A_R. 
		\end{align*} In particular, it also implies  $A(Y)\subseteq A_R$.  Furthermore,  by the fact that $A_R$ is star-shaped at  $A(Y)$,  we can get that   $$\text{conv}(A(Z)\cup A(Y))\subseteq A_R, ~~\forall Z\in A_{R}.$$
		
		Finally,  let $$\{A_\lambda:\lambda\in\Lambda\}:=\{{\rm conv}(A(Z)\cup A(Y)): Z\in A_R\}.$$ Clearly, $A_R=\cup_{\lambda\in \Lambda} A_\lambda$ and $Y\in B:=\bigcap_{\lambda\in \Lambda} A_\lambda\neq \emptyset$ and $R(X)=\bigcup_{\lambda \in\Lambda}R_{A_\lambda}(X)$ for all $X\in L_d^p$.  This is  what we are asking for.  
	\end{proof}
	
	The following proposition shows that a translation away from a set-valued risk measure can induce a star-shaped risk measure  under the correct choice of acceptance sets.
	
	\begin{proposition}\label{pro5.1}
		Suppose  $\{A_\lambda:\lambda\in\Lambda\}$ is a family of convex acceptance sets with  $A:=\bigcup_{\lambda\in\Lambda} A_\lambda$ and $B:=\bigcap_{\lambda\in\Lambda}  A_\lambda\neq \emptyset$.   Then,  for any $Y\in B$, the mapping  $R_Y: L_d^p\rightarrow\mathbb{F}_M$ defined as $$R_Y(X):= R_A(X+Y), ~~\forall X \in L_d^p,$$ is a set-valued star-shaped  risk measure.  In particular,  if $R_B(0)\neq \emptyset$, then for each $u\in R_B(0)$, $R(X):= R_A(X)-u$, $\forall X\in L^{p}_{d}$,  is a star-shaped risk measure.
	\end{proposition}
	\begin{proof}
		We claim that $A$ is star-shaped at $B$.  In fact,  for any $X\in A$, there exist $\lambda^*\in\Lambda$ such that $X\in A_{\lambda^*}$. For any $Y\in B$, $Y$ is also in $A_{\lambda^*}$.  Hence,  by the convexity of $A_{\lambda^*}$, it yields that $tX+(1-t)Y\in A_{\lambda^*}\subseteq A$ for all $t\in[0,1]$.
		
		Since $A$ satisfies (A1) and (A2), then for any $Y\in B$,  $R_{Y}(\cdot)$ is obvious a set-valued risk measure.  Denote  $A_Y$ as the acceptance set of $R_{Y}(\cdot)$.  Moreover, 
		\begin{align}\label{eq:ay}
			A_Y&=\{X\in L_d^p: 0\in R_Y(X)\}\nonumber\\
			&=\{X\in L_d^p: 0\in R_A(X+Y)\}\nonumber\\
			&=\{X\in L_d^p: X+Y\in A\}\nonumber\\
			&=A-Y.
		\end{align}
		
		Next, we claim that for any $Y\in B$, $A_Y$ is star-shaped at $0$.  Indeed, for any $Y\in B$ and $X\in A_{Y}$,  by the fact \eqref{eq:ay}, it follows $X+Y\in A$. Since 
		$A$ is star-shaped at $Y$, then it implies that, for any $t\in [0,1]$, $$t(X +Y)+(1-t)Y = tX +Y \in A.$$ This is equivalent to $tX \in A_Y$, which is star-shapedness at 0. 
		
		On the other hand,  since $R_Y(\cdot)=R_{A_Y}(\cdot)$ and $A_Y$ is star-shaped,  we conclude that $R_Y$ is star-shaped by Proposition \ref{pro3}.
		
		To see that $R$ is star-shaped, first note that $$R_B(0) = \{u\in M: u\in B\}=M\cap B\subseteq B.$$Then, we can choose $u\in R_B(0)\subseteq B$, then by the first part result of this proposition, we can get that  $R_{u}$ is a star-shaped set-valued risk measure.  More precisely,  $$R_{u}(X)= R_A(X+u)= R_A(X)-u=R(X), ~~\forall X\in L^{p}_{d},$$ thus $R$ is star-shaped.
	\end{proof}
		
	The following theorem unifies the  above discussions by exposing the interplay between set-valued risk measures and set-valued star-shaped risk measures.
	
	\begin{theorem}\label{theoguanxi}
		For a risk measure $R : L_d^p\to \mathbb{F}_M$, the following statements are equivalent. 
		
		(i) There exists $Y\in L_d^p$ such that $R_Y: L_d^p\rightarrow\mathbb{R}$ defined as,  $R_Y(X):=R(X+Y)$, $\forall X\in L^{p}_{d}$,  is a star-shaped risk measure.
		
		(ii) There exists a family of set-valued convex risk measures $\{R_\lambda:\lambda\in\Lambda\}$  such that $\bigcap_{\lambda\in\Lambda} A_{R_\lambda}\neq\emptyset$ and
		\begin{align}\label{eq:5.2}R(X)=\bigcup_{\lambda \in\Lambda}R_{\lambda}(X), ~~\forall X\in L_d^p.\end{align}
		In this case, we can take any $Y\in \bigcap_{\lambda\in\Lambda} A_{R_\lambda}$.
	\end{theorem}
	\begin{proof}
		(i) implies (ii).  Since $R_Y$ is star-shaped, then we have that $$A_{R_Y}=\{X\in L_d^p: 0\in R_Y(X)\}=\{X\in L_d^p: X+Y\in A_R\}=A_R-Y.$$ By the star-shapedness of  $A_{R_Y}$,  it then implies that $A_R$ is star-shaped at $Y$.  By Proposition \ref{pro5.2},  there exists a family of convex acceptance sets $\{A_\lambda:\lambda\in\Lambda\}:=\{\text{conv}(A(Z)\cup A(Y)):  Z\in A_R\}$ such that $A_R=\bigcup_{\lambda\in \Lambda} A_\lambda$, $\bigcap_{\lambda\in \Lambda} A_\lambda\neq \emptyset$ and 
		$$R(X)=\bigcup_{\lambda \in\Lambda}R_{A_\lambda}(X), ~\forall X\in L_d^p.$$
		Let $R_\lambda:=R_{A_\lambda}$, then $\{R_\lambda:\lambda\in\Lambda\}$ is the desired set-valued convex risk measures such that $\bigcap_{\lambda\in\Lambda} A_{R_\lambda}\neq\emptyset$ and \eqref{eq:5.2} holds. 
		
		(ii) implies (i). First note that  $R_Y$ is a set-valued risk measure, and we shall show that $R_Y$ is star-shaped.  This is obtained from \eqref{eq:5.2} and Proposition \ref{pro5.1}.
	\end{proof}
	
	\begin{remark}\label{remark5.2}
		In scalar-valued case,  for a given risk measure $\rho$, Proposition 6 in \cite{2022montaryandstarshaped} gave that for a family of acceptances $\Lambda$ and $B:= \bigcap_{\lambda\in \Lambda}A_\lambda$, $B\bigcap\mathbb{R}\ne\emptyset$ iff $B\ne\emptyset$ iff $\sup_{\lambda\in \Lambda}\rho_\lambda(0)=\rho_B(0)<\infty$.  If this equivalence relationship still holds for set-valued case, $\bigcap_{\lambda\in\Lambda} A_{R_\lambda}\neq\emptyset$ in Theorem \ref{theoguanxi} can be replaced by $R_B(0)=\bigcap_{\lambda\in\Lambda}R_\lambda(0)\ne\emptyset$. 
		
		However, in set-valued condition, the above equivalence condition no longer holds, i.e. $B\ne\emptyset$ is not enough to guarantee
		$R_B(0)=B\bigcap M\ne\emptyset$. For example, let $d=2$, $M=\mathbb{R}\times \{0\}$ and $K=\{(x_1,x_2):x_1\geq-x_2, x_2\geq 0\}$. For $B=(1,1)+K$, $B\ne\emptyset$, but $R_B(0)=B\bigcap M=\emptyset$ since $x_2\geq 1$ for all $(x_1,x_2)\in B$.
		
		Meanwhile, under certain conditions, this translation result in set-valued case can be unified with scalar-valued case, see the following proposition.
	\end{remark}
	\begin{proposition}\label{pro:5.3}
		Let $p=\infty$ and $M=\mathbb{R}^d$. Suppose $\{A_\lambda:\lambda\in\Lambda\}$ is a family of acceptance sets with $B:=\bigcap_{\lambda\in\Lambda}  A_\lambda\neq \emptyset$. Then $B\ne\emptyset$ if and only if $M\cap B\ne\emptyset$ if and only if
		$$\bigcap_{\lambda\in\Lambda} R_{A_\lambda}(0)=R_B(0)\ne\emptyset.$$
	\end{proposition}
	\begin{proof}
		Since we are on $L_d^\infty$ and the sets in $\{A_\lambda:\lambda\in\Lambda\}$ are monotone, $X\in B$ implies $\esssup X\in B$. In other words, $B\ne\emptyset$ if and only if $M\cap B\ne\emptyset$. Moreover, by 
		$$\bigcap_{\lambda\in\Lambda} R_{A_\lambda}(0)=\bigcap_{\lambda\in\Lambda}(M\cap A_\lambda)=M\cap B=R_B(0),$$
		the remaining is proved.
	\end{proof}
	
	The following corollary  can be directly obtained  from the above Theorem \ref{theoguanxi} and Proposition \ref{pro:5.3}. 
	\begin{corollary}
		For $M=\mathbb{R}^d$ and for a risk measure $R : L_d^\infty\to \mathbb{F}_M$, the following statements are equivalent. 
		
		(i) There exists $Y\in L_d^\infty$ such that $R_Y: L_d^\infty\rightarrow\mathbb{R}$ defined as,  $R_Y(X):=R(X+Y)$, $\forall X\in L^{\infty}_{d}$,  is a star-shaped risk measure.
		
		(ii) There exists a family of set-valued convex risk measures $\{R_\lambda:\lambda\in\Lambda\}$  such that $\bigcap_{\lambda\in\Lambda} R_\lambda(0)\neq\emptyset$ and
		\begin{align}R(X)=\bigcup_{\lambda \in\Lambda}R_{\lambda}(X), ~~\forall X\in L_d^\infty.\end{align}
		In this case, we can take any $Y\in \bigcap_{\lambda\in\Lambda} R_\lambda(0)\neq\emptyset$.
		
	\end{corollary}
	
	We end this section by an example of translating a set-valued risk measure to a set-valued star-shaped risk measure. 
	
	\begin{example}
		Let $R_\lambda(X):=R(X)+f(\lambda), \lambda\in\mathbb{R}$, where $R:L_d^p\to \mathbb{F}_M$ is a convex normalized risk measure and $f:\mathbb{R}\to\mathbb{R}\times\{0\}^{d-1}$ satisfies $\min_{\lambda\in \mathbb{R}}f(\lambda)=f(\epsilon)> -u$,  $\epsilon\in \mathbb{R}$ for any $u \in R(0)$. Note that $R_\lambda$ is not normalized, while it is clearly a convex risk measure. 
		
		We claim that $\cup_{\lambda\in \mathbb{R}}R_\lambda$ is not star-shaped. Indeed, since
		$$\bigcup_{\lambda\in \mathbb{R}}R_\lambda(X)=\bigcup_{\lambda\in \mathbb{R}}(R(X)+f(\lambda))=R(X)+\min_{\lambda\in \mathbb{R}} f(\lambda)=R(X)+f(\epsilon)=R_\epsilon(X),$$
		then for any $k>1$ and constant $X$, we have 
		$$R_\epsilon(kX)=R(kX)+f(\epsilon)=R(0)-kX+f(\epsilon),$$
		$$kR_\epsilon(X)=kR(X)+kf(\epsilon)=kR(0)-kX+kf(\epsilon).$$	 
		
		Since $f(\epsilon)+u> 0$ for any $u \in R(0)$,  and $R(0)+f(\epsilon)\nsubseteq k(R(0)+f(\epsilon))$. Hence, there is $X\in L_d^p$ such that $R_\epsilon(kX)\nsubseteq kR_\epsilon(X)$ for $k>1$, which implies that $R_\epsilon$ is not a star-shaped risk measure.
		
		However, note that this risk measure is only a translation away from a star-shaped risk measure. In fact, the functional
		
		$$R^*(X):=\bigcup_{\lambda\in \mathbb{R}}R_\lambda(X)-f(\epsilon)=R(X), ~~X\in L_d^p,$$
		is normalized and convex, thus star-shaped.		
	\end{example}	
	
	
\section{Conclusion} \label{sec6}
	In this paper, we introduce non-convex set-valued star-shaped risk measures. The representation theorems of set-valued risk measures and the set-valued normalized star-shaped risk measure are given, and the relationship between set-valued risk measures and set-valued star-shaped risk measures is established. It is also interesting to investigate the corresponding  set-valued star-shaped risk measures in dynamic framework. The readers can refer to \cite{2013timeconsistency}, \cite{2023amw} or \cite{2021af} for dynamic set-valued risk measures.   We leave it for the future research. 
 
    
	\section*{Acknowledgments}  
The authors would like to thank the editors and two referees for their valuable comments and suggestions which led to a much improved version of the paper. The authors thank the grant supported by the Fundamental Research Funds for the Central Universities (No. 2024KYJD2008). 
	
	
	\bibliographystyle{jf}

\end{document}